# Realism of Simulation Models in Serious Gaming

## Two case studies from Urban Water Management Higher Education

Darwin Droll[1] and Heinrich Söbke[1] [0000-0002-0105-3126]

[2] Bauhaus-Universität Weimar, Bauhaus-Institute for Infrastructure Solutions (b.is), Goetheplatz 7/8, 99423 Weimar, Germany

{Darwin-droll@web.de| heinrich.soebke@uni-weimar.de}

**Abstract.** For games used in educational contexts, realism, i.e., the degree of congruence between the simulation models used in the games and the real-world systems represented, is an important characteristic for achieving learning goals well. However, in the past, the realism of especially entertainment games has often been identified as insufficient. Thus, this study is investigating the degree of realism provided by current games. To this purpose, two games in the domain urban water management, a subdomain of environmental engineering (EE), are examined. One is ANAWAK, a web-based serious game on water management and climate change. For ANAWAK, an analysis of the simulation model is conducted. Second, the simulation model of the entertainment game Cities: Skylines (CS) is analyzed. In addition, a survey among CS players (N=61) is conducted. Thereby, different degrees of realism in various EE subdomains are revealed. All in all, there are still considerable deficits regarding the degree of realism in the CS simulation model. However, modding as a means of achieving more realistic simulation models is more widely supported than in the past.

## 1       Introduction

The domain environmental engineering (EE) addresses the engineered design of the environment and includes several subdomains, such as urban water management, waste management, energy, and transportation. Among key engineering design tools are simulations [1]. Simulation refers to the time-dependent imitation of a real-world system [2]. The underlying simulation model is a purpose-oriented abstraction of the real-world system to be simulated, i.e., only those aspects of the system that are relevant to the purpose are modeled.

**Serious Games.** Simulations are also applied effectively in educational contexts of engineering domains [3], for example to enable learners to experiment exploratively with simulation models and thus to understand the systems simulated, hence the learning content. Some of the simulations in educational contexts take the form of digital games. Fundamental part of digital games are simulation models. The games aim to motivate learners and thus increase the learners' engagement with the games and thus indirectly with the learning content [4]. For example, in the EE subdomain urban water management, exist many educational simulation games, also referred to as serious games [5–8].

**Realism.** The domain-specific content of games is considered very relevant for learning [9]: domain-specific content should contribute to the authenticity of the learning environment [10], and consists, among other things, of the terminology and simulation models used. Furthermore, the domain-specific content of serious games is supposed to be as close to reality as possible [11, 12]. In this regard, simulation models are considered to have a driving role in serious games [13, 14]. Tashiro and Dunlap [15] regard realism as one of the key dimensions of learning outcomes in educational games, along with engagement. A term used similarly to authenticity of domain-specific content is validity [13]. In serious games with the purpose of learning, one may also refer to learning goal orientation, i.e., the aspects of the systems modelled that are necessary to achieve the learning goals are considered specifically in the simulation models. Accordingly, realism might be defined as the degree of congruence between the simulation models used in the games and the real-world systems to be represented. Thus, in this study, the term realism is used to describe the quality of games whose domain-specific content in the form of simulation models and terminology is oriented as closely as possible to the relevant aspects of the modeled system being addressed in the game. Not considered in this study is visual realism, which is beneficial for the acceptance of what is presented only up to a certain limit [16]. In the literature, this proposition is also known as *Uncanny Valley* [17], which refers to the declining acceptance of artificial characters above a certain level of realism.

**Entertainment Games as Serious Games.** In addition to serious games created explicitly for learning purposes, the use of entertainment games in educational contexts is also common, such as *Civilization* [18] or *Roller Coster Tycoon* [19]. Among the advantages of using entertainment games as serious games is the non-existent development effort [20]. At the same time, entertainment games are optimized for being enjoyable, i.e., entertainment games are characterized by high attractiveness. However, even in the case of entertainment games that are used for learning, the game domain-specific content—to a large extent the simulation models— needs to be oriented to learning goals, i.e., the games show a high degree of realism in the previously defined sense. However, in general, simulation models of entertainment games in educational contexts are attributed inaccuracies [21]. The entertainment game studied here, Cities: Skylines, also appears to lack technical accuracy for the domain urban planning [22]. Such inaccuracies of simulation models might need to be considered in the design of learning scenarios, for example, for addressing learning outcomes outside of the game [23]. Thus, the value of entertainment games is also often seen in further effects, e.g., as a trigger for learning processes with further media [24]. For example, Bawa et al. [25] exploit the motivation nurtured from MMOGs for learning with media beyond the MMOG, such as wikis.

**The Case of SimCity.** The impetus for the study outlined in this article arose from the observation that entertainment games are not necessarily designed with learning goals in mind. More precisely, entertainment games often do not offer the degree of realism necessary to support specific learning goals. This may also be the case with entertainment games that appear to be very realistic at first glance. The case that illustrates these statements is SimCity [26] in relation to the learning objectives of the domain urban water management [27].

In consequence, exemplarily from the perspective of the domain urban water management, thus, the research question arises to what extent the progress in game development in recent years, visible among others in the currently most popular city builder game Cities: Skylines (CS) [28], has led to a higher degree of realism, which is expressed for example by more detailed simulation models that might support a wider range of learning goals than before. This research question yet recognizes that by carefully aligning learning goals and learning scenario using an entertainment game, some domain-specific learning goals may be achievable, and that beyond domain-specific learning objectives, additional higher-order learning goals (e.g., metacognitive learning goals) may be reachable. However, the research questions targets the extent, to which domain-specific learning objectives may be achieved without significant alignment effort.

The remainder of the paper is organized as follows: in the next section the methodology applied is outlined and the section thereafter describes the two games. Section 4 presents the results, which are discussed in Section 5 along with the limitations of this study. Finally, section 6 summarizes the findings.

## 2    Method

For an evaluation of the current degree of realism of simulation games in urban water management, the simulation models of two simulation games were analyzed. The selection of the games was guided by representing each of the two poles—, specifically developed serious game and entertainment game—, a specific educational scope was not required, nevertheless, both games are suitable for graduate education use. Therefore, sufficient complexity of the simulation models is reasonable allowing an assessment of the degree of realism provided by the games. One game was a specifically developed serious game (ANAWAK), which was analyzed for providing a basis for comparison. Second, the components related to urban water management in CS were analyzed. In each case, the model analysis was performed by two EE experts. For the analysis of CS, both experts each spent about 20 h of playing time, while ANAWAK's lower complexity required only about 5 h each. Both first independently assessed the games for urban water management components, such as water treatment plants (WTPs), identified parameters of the components selected, and analyzed whether and how they were presumed to be incorporated into the simulation model. In cases where the experts' assessments did not correspond, the experts elaborated in discussions unified assessments. Only in-game observations and evaluation of documentation were exploited for the model analyses; source code or other formal models were not available. The model analyses were supplemented by an online survey of CS players to collect players' assessment of the degree of realism of CS simulation models with respect to EE systems.

## 3    Game Descriptions

**ANAWAK.** ANAWAK is a serious game designed for learning purposes [29]. ANAWAK is the result of a research project on climate adaptation [30]. The learning content of the game originates from the research project and relates to the basics of urban water management in times of climate change. The game, which is unfortunately currently offline, is a web-based game asking the player to take over the water management of a settlement, and to make turn-based decisions that affect the settlement's water supply. On a simplified terrain view (Fig. 1), the player interactively influences various game components. Current and past states of the game with respect to the subgoals of sustainability assessment (ecology, economy, and society) are available through a menu. At the end of the game, which takes about 20 minutes to play, the result is presented. ANAWAK was chosen as representative for a group of serious games which, departing from the learning goals, were implemented with rather unelaborate aesthetics and rather simple game mechanics and thus, — remarkably — according to the above definition, show a high degree of realism.

Fig. 1. Game scene ANAWAK (Translations added by authors)

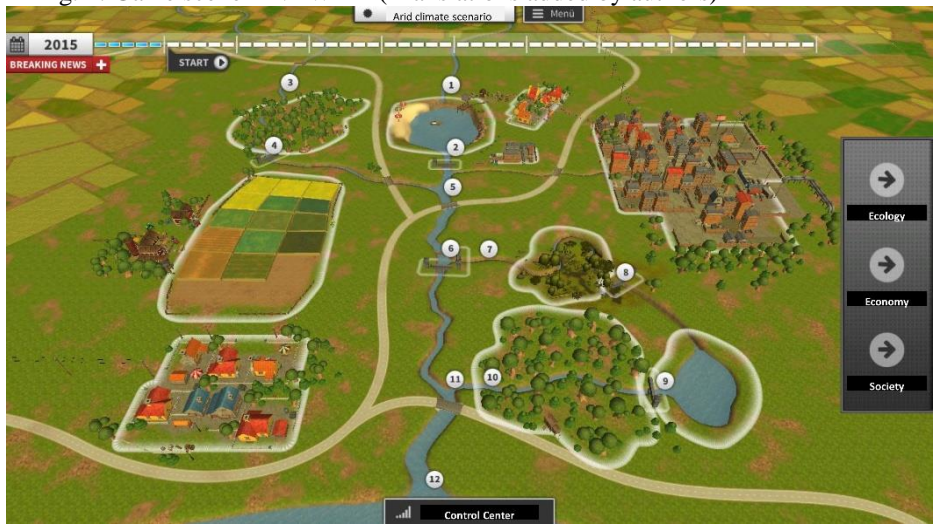

**Cities: Skylines (CS).** Since its launch in 2015, CS has been one of the most popular city builder games, which is reputed having a high degree of realism and has also been used in real urban planning contexts, e.g., in transportation planning for the city of Stockholm [31]. CS exhibits a high degree of realism especially in the domain transportation [32]. Actual georeferenced data have also been displayed in CS [33]. Overall, therefore, as a recent game with detailed simulation models, CS appeared to be a candidate for investigating the degree of realism of urban water management domain-specific simulation models in an entertainment city builder game, especially since CS is considered a de facto successor to SimCity.

## 4 Results

### 4.1 Simulation Model Analysis of ANAWAK

ANAWAK employs a static component arrangement, i.e., players cannot change the structure of the game scenario. The simulation model of ANAWAK is described with a qualitative representation of how subgoals are influenced by parameters of components. Taken from the game documentation, Table 1 provides a list of urban water management-related components that may be manipulated by players and are controlled by parameters (shown indented). Adjusting the parameters changes the status of the components and affects the subgoals. For example, for the urban drainage component, increasing the amount of swale infiltration has a positive effect on ecology, economy, and society, respectively. In addition to the components shown in the table, there are five weirs usable for controlling water distribution. However, no consistent interdependencies with the assessment subgoals are definable. The dependencies shown in Table 1 define a significant part of the learning goals, for example high water demand (Table 1: 3) is considered negative for the subgoal ecology. The results of the game actions may be gauged from the values of the subgoals on the one hand, but also from the values of read-only parameters of the components on the other hand. Among the components that exclusively feature such read-only parameters are two lakes, whose filling level and water quality may be observed, a forest (parameter: groundwater level) and a wetland (parameter: water quality). Experimentation with the model (manipulation of parameters and subsequent observation of interdependencies) accounts for learning.

### 4.2 Simulation Model Analysis of Cities: Skylines

Since CS is a city builder game, players build cities from components and thus influence the simulation model. CS was examined for components relevant to urban water management based on the authors' gaming experiences and on developer-provided documentation [34]. The following enumeration describes the components identified and introduces discrepancies to reality using the keyword *Reality Check*.

- **Pipes.** Water pipes supply water and discharge wastewater. *Reality Check*: Drinking water pipes and wastewater pipes are always positioned simultaneously and therefore on the same route, which is independent of further construction, such as roads. Parameters such as flow rates, pressure or pipe gradients do not exist.
- **Water Tower**. The water tower acts as a groundwater pump, providing a constant flow of 60,000 m³/week. If water tower is constructed in an industrial area, polluted water is pumped. *Reality Check*: There is no connection with precipitation water, groundwater is assumed to be an infinite reservoir.

- **Water Pumping Station**. A water pumping station conveys with constant flow of 120,000 m³/week, is placed at a water body and connected to the water network. *Reality Check:* The capacity of the water body is rather overestimated.

- **Water Drain Pipe.** The water drain pipe is initially the only means of discharging wastewater, it is connected to a water body and discharges untreated wastewater with a flow of 120,000 m³/week into the water body. The pollution is visualized after a short time by brown coloration of the water body in the flow direction below the wastewater pipe. The water pumping station therefore is to be built in flow direction above the water drain pipe, otherwise polluted water will be pumped. *Reality Check*: Water drain pipes are not very common in the western industrialized countries simulated in CS.

- **Water Treatment Plant.** The water treatment plant is unlocked after progressing in the game and cleans a wastewater flow of 16,000 m³/week at 85%. *Reality Check*: No distinction is made between purification process steps, further, the 85% purification performance is given in general and not on specific contaminants.

- **Rain.** CS features rain. *Reality Check:* Rain has no influence on water management components such as water bodies or groundwater.

|  |  | Ecology | Economy | Society |
|---|---|---|---|---|
| **a)** | **Agriculture** |  |  |  |
|  | **(1) Share of arable land** | - | + | + |
|  | **(2) Share of corn in arable land** | --- | +++ | - |
| **b)** | **Auendorf (settlement)** |  |  |  |
|  | **(3) Water demand** | - | + | ++ |
| **c)** | **Urban drainage** |  |  |  |
|  | **(4) Rain retention basin** | ++ | - | + |
|  | **(5) Green roofs** | ++ | - | ++ |
|  | **(6) Swale infiltration** | + | + | + |
| **d)** | **Nordwald (forest)** |  |  |  |
|  | **(7) Share of deciduous forest** | +++ | + | ++ |
| **e)** | **Waterworks** |  |  |  |
|  | **(8) Water demand bank filtrate** | -- | + | ++ |

Table 1. ANAWAK simulation model: qualitative interdependencies (+: correlation in the same direction, -: correlation in the opposite direction; number of + and -: strength of correlation).

**CS extensions.** A prominent feature of CS is its support for modding. The software is open for programmatic program changes and is further developed by the developer ("extension packs") as well as by third parties ("mods"). Both, certain extension packs and several mods complement CS from the perspective of urban water management. The following are examples:

- **Heating Pipes.** [35] The heating pipes are part of the expansion pack *Cities: Snowfall*, which adds heating pipes to the water network and interlinks energy management and urban water management.
- **Stabilization Basin.** [36] The stabilization basin is an alternative to the water treatment plant. The stabilization basin percolates the water and therefore has a low capacity only.
- **Ground Water Pump.** [37] The pump operates independently of a water body and differs from the water tower not functionally, but in appearance and parameters.
- **Modular Sewage Treatment Facility.** [38] Very close to real water treatment plants are the buildings of this mod, which represent all the main steps of the purification process and is modeled after a real water treatment plant in Denmark. *Reality Check*: Again, no specific contaminants in the water are distinguished.

Basically, mods and expansion packs provide for a continuous expansion of the CS simulation model, so that currently also significantly more types of water treatment plants are available. In this respect, progress can be seen compared to SimCity. However, the expansions refer to the components themselves, but the descriptive parameters of the components remain unchanged and are the same for all components, such as cost, upkeep, noise pollution, water, electricity consumption and size. This limitation renders developing domain-specific extensions harder.

### 4.3 Questionnaire: CS as a Serious Game in EE

The questionnaire was announced as a survey on CS as a serious game in EE and was available in June 2018 in CS online communities accessible through the homepage of the publisher Paradox. Of the 84 responses, 61 were evaluable. All Likert scales used had 7 points.

**Demographics. Sex**: 91% of participants identified themselves as male, 9% as female. **Age**: 39% of participants were older than 30 years, 17% were in the 25-30 age group, 19% were 19-24, 24% were 12-18, and 1% were younger than 12. **Occupation**: 25% were high school students, 20% were college students, 30% were employed and 25 chose *Other*. **Playing Time**: on average, participants played CS for 3.6 h (SD = 1.77) per week. **EE Affinity**: 44% denied an interest in EE. 30% affirmed this question in general and 26% specifically since engaging with CS.

**Realism.** To the question "How realistic is CS in terms of accurate representation of real systems?" the participants answered with an average of 4.4 points (SD= 1.43). Participants were also asked about the degree of realism in CS of different subdomains of EE. *Transportation* (M=4.9, SD=1.46) was rated by far the most realistic. Waste Management (M=4.9, SD=1.46), *Wastewater Discharge* (M=3.4, SD=1.13), *Energy* (M=3.3, SD=1.08) and *Water Supply* (M=3.1, SD=1.14), both subdomains of urban water management, tended to be perceived less realistic. With respect to individual buildings (Fig. 2), the participants gave widely differing ratings from M=5.5 to M=3.9: the most realistic rating was given to the *Wind Turbine*, which, however, is also not very complex in reality and hence may be easily simulated. The *Airport* was rated least realistic, a component that is, however, highly complex in reality. In addition, the participants were also asked in an open question about the differences between the buildings shown in CS and the corresponding real buildings. As an example, for the water treatment plant (WTP), a total of ten responses were given. While one answer considered the CS WTP to be very realistic, as it reduces pollution strongly, but not down to drinking water quality, other answers rather saw the differences: sewage is only treated and dumped in the CS WTP, while in the real WTP there might be a complete recycling. There are also different types of WTPs in reality not being available in CS. Further answers point to a too low cleaning performance (4 answers), a too simple simulation model (3 answers) and a quantitatively inaccurate simulation model (2 answers).

**Enjoyable Aspects.** In a further question, the participants were asked to indicate the importance of various aspects for the game enjoyment on a Likert scale. *Realism* ranked in the middle with M=4.3 (SD=1.31). Placed ahead were *City Building* (M=5.4, SD=0.90) and *Space for Creativity* (M=5.2, SD=0.80), each important motivations for playing a city builder game like CS. Less important, but still above the scale's mid-point, were *Challenge* (M=3.8, SD=1.38) and *Online Community* (M=3.5, SD=1.79).

<p align="center">Fig. 2 Degree of realism of buildings<br/>(7-point Likert scale, option "I do not know"; N ranged from 38 to 58)</p>

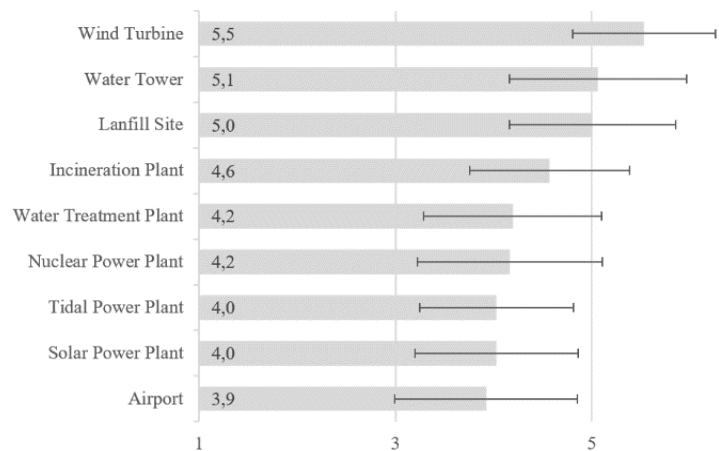

**Attitudes towards Digital Games as Learning Tools.** The question to what extent participants believe that computer games are generally useful as learning tools scored M=6.3 points (SD=1.00). The question about the extent to which CS is seen as a learning tool scored M=4.3 points (SD=1.34) for the domain EE and M=5.6 (SD=1.34) for other domains in general. EE tends to be seen as less appropriate as other domains, such as urban planning, which was mentioned multiple times in the comments. Table 2 shows the responses to the multiple-choice question about which conceptual statements are taught by playing CS. Again, different emphases in CS are evident.

| Conceptual statement | Frequency |
|---|---|
| Land Fill Sites Are Not Sustainable | 74.6 % |
| Wind Energy Is Unsteady | 47.6 % |
| Recycling | 44.4 % |
| Green Power Is Expensive | 44.4 % |
| Water Resources are Not Infinite | 33.3 % |
| Water Pipes Should Mainly Placed in Streets | 28.6 % |
| Storage of Energy | 15.9 % |
| Nuclear Waste Is a Problem | 11.1 % |

Table 2. Assessment of the extent to which CS contributes to a basic understanding of various concepts of EE (multiple responses allowed, N=61).

**Mods.** 79% of the participants indicated using mods. The reasons for using mods were asked in a multiple selection question. A significant proportion (40%) of participants install mods to improve the accuracy of simulation models. These high numbers might be motivated by participants playing CS rather frequently, and on the other hand by participants also being rather serious game-affine. Moreover, 94% would like to improve the gameplay, 64% would like to increase the choice of buildings, 32% would like to increase the quality of the graphics, and 8% have other reasons for using mods.

## 5    Discussion

A primary limitation of the study is that it is based on only two games and thus is not representative. Nevertheless, it is to be assumed—and this is aligned with the literature referenced in the introduction—that in most serious games specifically developed, the simulation models show a similar learning goal-oriented approach, and that in most other entertainment games the degree of realism is similarly simplistic. Likewise, there is the restriction that the analysis of CS refers to the initial version without extension packs. However, even a subsequent ad-hoc analysis of the extension packs released to date did not reveal any qualitative discrepancies from the results.

The results of the survey are subject to the caveat that the assessments of realism were not given by experts but by CS players, who might lack the necessary knowledge to provide profound assessments. However, for example, the assessment of the domain transportation as the most realistic is aligned with the literature [32]. Further, in specific assessments, such as the assessment of the realism of buildings, participants could deny an assessment using the option "I do not know". Moreover, the assessments expressed in comments indicated participants in the survey to be quite reflective.

Compared to SimCity, CS does not show any significant progress in the level of detail of the simulation model from an EE perspective. However, due to the game engine used [39], CS offers the possibility to comprehensively access the simulation model via modding. Therefore, building a simulation model with higher degrees of realism is costly, but in general conceivable. The necessary degree of realism depends on the learning goals. It may be sufficient to support qualitative learning goals, as is the case in ANAWAK. However, with the appropriate level of detail in the simulation model, supporting quantitative learning goals is also feasible, albeit effortful.

## 6    Conclusions

More than a decade ago, it has been reported that the simulation model of the entertainment game SimCity is optimized for entertainment purposes rather than providing environmental engineering (EE) specific knowledge. Hereby, potential learning goals supported by SimCity in learning contexts are limited. Thus, the aim of this study was to conduct a current review of realism in entertainment games regarding the EE subdomain urban water management. Simulation models of two games, the dedicated serious game ANAWAK and the popular entertainment game Cities: Skylines (CS) were analyzed in relation to urban water management. While ANAWAK is an example of aligning simulation models with learning goals in a simple interactive graphical presentation, the simulation model of CS shows little progress compared to the findings more than a decade ago with regard to realism of domain-specific content. In contrast, progress is evident in the support of extensions to be developed individually, so-called mods, which may be a path to more realism of simulation models. In addition to the model analysis, an online questionnaire was employed to elicit feedback from serious game interested members of the CS community on how realism of the simulation model in CS is perceived in relation to EE and to what extent CS is considered to be an effective learning tool. Findings include that

- the realism of the simulations model regarding various EE subdomain differs widely,
- players are aware of the limitations of simulation models, and
- various domain-specific concepts are nevertheless taught by CS.

Overall, the findings indicate that–when using entertainment games for conveying domain knowledge–matching the learning goals to the games' simulation models is still required. Technological advances have not resulted in simulation models showing significantly more realism, but a more purposeful development of sophisticated mods is supported. Although we have no indication that these findings are limited to the domain of urban water management, the generalizability of the findings to other domains remains to be confirmed through further research.

## Acknowledgements

The authors gratefully acknowledge the financial support provided by the German Federal Ministry of Education and Research (BMBF) through grant FKZ 033W011B provided for the "TWIST++" project Any opinions, findings, conclusions, or recommendations expressed in this paper are those of the authors and do not necessarily reflect the views of the institution mentioned above.